\documentclass[twocolumn,aps,showpacs]{revtex4}

\begin{document}

\title{Particle Statistics in Quantum Information
Processing}

\author{Yasser Omar}

\affiliation{Centro de F\'isica de Plasmas\\Instituto Superior
T\'ecnico, P-1049-001 Lisbon, Portugal}

\date{30 August 2004}

\begin{abstract}
Particle statistics is a fundamental part of quantum physics, and
yet its role and use in the context of quantum information have
been poorly explored so far. After briefly introducing particle
statistics and the Symmetrization Postulate, I will argue that
this fundamental aspect of Nature can be seen as a resource for
quantum information processing and I will present examples showing
how it is possible to do useful and efficient quantum information
processing using only the effects of particles statistics.
\end{abstract}

\pacs{05.30.-d, 03.67.-a}

\maketitle

\section{An Extra Postulate is Required}

I believe most physicists would consider that the postulates (or
at least the properties they embody) concerning the superposition,
evolution and measurement of quantum states cover the essence of
Quantum Mechanics, the theory that is at the basis of current
fundamental Physics and gives us such an accurate description of
Nature at the atomic scale. Yet, if the theory was only based on
these postulates (or properties), its descriptive power would be
almost zero and its interest, if any, would be mainly
mathematical. As soon as one wants to describe matter, one has to
include an extra postulate: \emph{Pauli's Exclusion Principle}.
One of its usual formulations, equivalent to the one proposed
originally by Wolfang Pauli in 1925 \cite{pauli-exclusion}, is the
following:\\

\textbf{Pauli's Exclusion Principle} --- \emph{No two electrons
can share the same quantum numbers.}\\

This principle refers to electrons, which constitute a significant
(but not the whole) part of matter, and is crucial in helping us
explain a wide range of phenomena, including:

\begin{itemize}

\item The electronic structure of atoms and, as a consequence, the
whole Periodic Table;

\item The electronic structure of solids and their electrical and
thermal pro\-perties;

\item The formation of white dwarfs, where the gravitational
collapse of the star is halted by the pressure resulting from its
electrons being unable to occupy the same states;

\item The repulsive force that is part of the \emph{ionic bond} of
molecules and puts a limit to how close the ions can get (e.g.,
$0.28\!$ nm between $Na^+$ and $Cl^-$ for solid sodium chloride),
given the restrictions to the states the overlapping electrons can
share.

\end{itemize}
We thus see how Pauli's insight when proposing the Exclusion
Principle was fundamental for the success of Quantum Mechanics.
Although he made many other important contributions to Physics, it
was for this one that he was awarded the Nobel prize in 1945.

Pauli's Exclusion Principle remains as a postulate, for Pauli's
own dissatisfaction, as he expressed in his Nobel prize acceptance
lecture in 1946 \cite{pauli-lecture}:

\begin{quote}
\emph{``Already in my original paper I stressed the circumstance
that I was unable to give a logical reason for the exclusion
principle or to deduce it from more general assumptions. I had
always the feeling, and I still have it today, that this is a
deficiency." }
\end{quote}
In any case, as inexplicable as it may be, Pauli's Exclusion
Principle seems to beg for a generalization. In fact, it was soon
realized that other particles apart from electrons suffer from the
same inability to share a common quantum state (e.g., protons).
More surprising was the indication that some particles seem to
obey to the exactly opposite effect, being
--- under certain circumstances --- forced to share a common state, as
for instance photons in the stimulated emission phenomenon, thus
calling for a much more drastic generalization of Pauli's
Principle.

\section{Identity and Indistinguishability}

Pauli's Exclusion Principle intervenes in a wide range of
phenomena, from the chemical bond in the salt on our table to the
formation of stars in distant galaxies. This is because it applies
to electrons and we consider all electrons in the universe to be
\emph{identical}, as well as any other kind of quantum
particles:\\

\textbf{Identical particles} --- \emph{Two particles are said to
be \emph{identical} if all their intrinsic properties (e.g., mass,
electrical charge, spin, colour, ...) are exactly the same.}\\

Thus, not only all electrons are identical, but also all
positrons, photons, protons, neutrons, up quarks, muon neutrinos,
hydrogen atoms, etc. They each have the same defining properties
and behave the same way under the interactions associated with
those properties. This brings us to yet another purely quantum
effect, that of \emph{indistinguishable} particles.

How can we distinguish identical particles? Their possibly
different internal states are not a good criterium, as the
dynamics can in general affect the internal degrees of freedom of
the particles. The same is valid for their momentum or other
dynamical variables. But their spatial location can actually be
used to distinguish them. Let us imagine we have two identical
particles, one in Alice's possession and the other with Bob. If
these two parties are kept distant enough so that the wave
functions of the particles practically never overlap (during the
time we consider this system), then it is possible to keep track
of the particles just by keeping track of the classical parties.
This situation is not uncommon in quantum mechanics. If, on the
other hand, the wave functions do overlap at some point, then we
no longer know which particle is with which party. And if we just
do not or cannot involve these classical parties at all, then it
is in general also impossible to keep track of identical
particles. In both these cases, the particles become completely
indistinguishable, they are identified by completely arbitrary
labels, with no physical meaning (as opposed to \emph{Alice} and
\emph{Bob}). In these situations the description of our system
becomes ambiguous and the so-called \textit{exchange degeneracy}
appears.
%
The problem of finding the correct and unambiguous description for
such systems is very general and requires the introduction of a
new
postulate for quantum mechanics: the Symmetrization Postulate.\\

\textbf{Symmetrization Postulate} \label{Post. Symmetrization} ---
\emph{In a system containing indistinguishable particles, the only
possible states of the system are the ones described by vectors
that are, with respect to permutations of the labels of those
particles:}
\begin{itemize}

\item \textit{either \emph{completely symmetrical} --- in which
case the particles are called \emph{bosons};}

\item \textit{either \emph{completely antisymmetrical} --- in
which case the particles are called \emph{fermions}.}

\end{itemize}

This is in fact a generalization of Pauli's Exclusion Principle,
in two ways. First, it extends it to a whole class of particles
which suffer the same restrictions: fermions. But if goes even
further and introduces a new class of particles, bosons, which
have a very different behaviour, almost the opposite, as they are
forced to share the same quantum numbers. To decide which
particles should be associated to a particular symmetry is
something that must ultimately be determined by observation. The
Symmetrization Postulate matches the study of such symmetries with
our empirical knowledge: as far as we know today, there are two
classes of particles in Nature according to their collective
behaviour in indistinguishable situations. These are, of course,
bosons and fermions: no particles have been found so far that
under the same circumstances could be described by vectors that
are neither symmetrical nor antisymmetrical. It is important to
note that none of this could have been deduced from the other
standard postulates of Quantum Mechanics. Yet, the Symmetrization
Postulate is rarely evoked.

\subsection{The Spin-Statistics Connection}

To determine whether a given particle is a fermion or a boson, we
need to investigate its statistical behaviour in the presence of
(at least one) other identical particles, when they are all
indistinguishable, and this behaviour will be very different for
the two types of particles. Indirect methods could also help us
reach a conclusion, but before any of that a simple and intriguing
property can actually come to our rescue: the
\emph{spin-statistics connection}.\\

\textbf{Spin-Statistics Theorem} --- \textit{Particles with
integer spin are bosons. Particles with half-integer spin are
fermions.}\\

This is not only a widely known empirical rule in Physics, but in
fact a theorem (originally proved by Pauli \cite{pauli-theorem}),
even if its proofs are not all completely clear and free from
controversy. Thanks to it, it is very easy to determine whether
some particle is either a fermion or a boson. In particular, this
criterion works also for composite particles. It is quite
surprising to find such a connection between the spin of a
particle and its statistical nature, a connection whose origins I
believe are still not well understood.

\section{Quantum Information}

The use of quantum systems and their unique properties to encode,
transmit, process and store information offers a completely new
way to deal with information, representing a revolution for
Information Sciences, and possibly for our Information Society as
well. It is conceivable that one day we will have a more
fundamental description of Nature than Quantum Physics and this
may well represent yet again a revolution in the way we deal with
information. But before trying to reach that far, we should ask
ourselves if we have already explored all the properties of the
quantum world in terms of their relevance for information
processing. I think not. There is still at least one other
property, as fundamental as the ones already mentioned, that
should be considered: particle statistics \footnote{Also referred
to as \emph{quantum statistics}; I shall use both expressions
interchangeably.}, or the apparent fact that every particle is
either a fermion or a boson and that their collective behaviour
obeys precise rules. Now, can the effects of particle statistics
play any role in quantum information processing? Can they be used
to perform useful quantum information tasks? And in an efficient
way?

For the last couple of years we have been exploring the role of
indistinguishable particles and quantum statistics in quantum
information processing, both for fermions and bosons \footnote{
Note also some recent attempts to use the statistical properties
of particles in the context of quantum information using electrons
\cite{divincenzo-loss, antonio}, photons \cite{dik}, parahydrogen
\cite{glaser}, fermions \cite{lloyd}, bosons \cite{sougato}, and
anyons \cite{kitaev}, but never presenting a systematic comparison
between the fermionic and bosonic statistics (This note and the
respective references had to be removed from the published version
of this article due to length restrictions).}. We have proved
that, using \emph{only} the effects of particle statistics, it is
possible to perform a quantum information task
--- such as transfer of entanglement \cite{Omar}, to do useful quantum
information processing --- such as entanglement concentration
\cite{Paunkovic}, and do it in a optimal way --- in particular, in
a state discrimination protocol \cite{Bose}. All these results
make use the antibunching of indistinguishable electrons impinging
in a beam splitter, as well as of the bunching of photons in a
similar situation, both a clear signature of their statistics
\cite{yamamoto}.

Using two pairs of entangled particles, it was shown for both
fermions (electrons) and bosons (photons) that
indistinguishability enforces a transfer of entanglement from the
internal to the spatial degrees of freedom without any interaction
between these degrees of freedom \cite{Omar}. Furthermore,
sub-ensembles selected by local measurements of the path will in
general have different amounts of entanglement in the internal
degrees of freedom depending on the statistics of the particles
involved. Then, an entanglement concentration scheme was proposed
which uses only the effects of particle statistics
\cite{Paunkovic}. Although its efficiency is the same for both
fermions and bosons, the protocol itself is slightly different
depending on the nature of the particles. Moreover, no explicit
controlled operation is required at any stage. Finally, particle
statistics is applied to the problem of optimal ambiguous
discrimination of quantum states \cite{Bose}. It was shown that
the Helstrom optimal single-shot discrimination probability to
distinguish non-orthogonal states of two qubits (encoded in the
internal degree of freedom of two electrons or two photons) can be
achieved using only the properties of fermions and bosons.
Furthermore, this method offers interesting applications to the
detection of entanglement and the purification of mixed states.

Two main features emerge from the above results: particle
statistics appears as a resource that can replace controlled
operations (conditional interactions) in a \emph{natural} way, and
information processing using indistinguishable particles is
different for fermions and bosons. The obtained results can also
be tested with current technology. Moreover, they establish that
indistinguishable particles and quantum statistics can play a new
and important role in quantum information and that this connection
should be further explored.

\section*{Acknowledgements}

This article is based on a talk delivered at the
\emph{International Meeting on Quantum Information Science:
Foundations of Quantum Information}, held in Camerino, Italy, in
April 2004. I would like to thank the support from
Funda\c{c}\~{a}o para a Ci\^{e}ncia e a Tecnologia (Portugal) and
the 3rd Community Support Framework of the European Social Fund,
as well as acknowledge the QuantLog initiative.

\end{document}